# Seasonal Variability of Pluto's Haze Formation Revealed by Laboratory Simulations

Zhengbo Yang[1], Chao He[1,*], Yu Liu[1,*], Sai Wang[1], Haixin Li[1], Yingjian Wang[1], Xiao'ou Luo[1], Sarah M. Hörst[2,3], Sarah E. Moran[4,3], Véronique Vuitton[5], Laurène Flandinet[5], Patricia McGuiggan[6]

1. School of Earth and Space Sciences, University of Science and Technology of China, Hefei 230026, China
2. Department of Earth and Planetary Sciences, Johns Hopkins University, Baltimore, MD, USA.
3. Space Telescope Science Institute, Baltimore, MD, USA.
4. NHFP Sagan Fellow, NASA Goddard Space Flight Center, Greenbelt, MD 20771, USA.
5. Univ. Grenoble Alpes, CNRS, IPAG, 38000 Grenoble, France.
6. Department of Materials Science and Engineering, Johns Hopkins University, Baltimore, MD, USA

Corresponding authors: Chao He (chaohe23@ustc.edu.cn), Yu Liu (yliu001@ustc.edu.cn)

## Abstract

Pluto possesses a thin atmosphere primarily composed of $N_2$, with minor constituents including CO and $CH_4$. Photochemical processes generate distinct haze layers as observed by the New Horizons spacecraft. However, the mechanisms governing haze formation, as well as the composition and physical properties of the hazes, remain poorly constrained. Due to Pluto's highly eccentric orbit and obliquity, its surface temperature and atmospheric composition undergo substantial seasonal variations, but it is unclear how such seasonal variations impact the chemical pathways and efficiency of haze formation in Pluto's atmosphere. To address this, we conducted a laboratory simulation of Pluto's atmospheric photochemistry, in which $N_2$/$CH_4$/CO gas mixtures with $CH_4$ concentrations varying from 0.1% to 5% were exposed to a glow discharge to initiate photochemical reactions. Gas-phase composition was monitored in situ using a residual gas analyzer (RGA), while the solid-phase products were characterized by atomic force microscopy (AFM), a gas pycnometer, infrared spectroscopy (IR), and very high-resolution mass spectrometry (VHRMS) to determine particle sizes, density, and composition, respectively. Our results show that increasing the $CH_4$ mixing ratio significantly enhances the yield of gas and solid products. Under low $CH_4$ conditions, nitrogen is primarily incorporated into solids as cyanide groups; whereas $CH_4$-rich conditions favor the formation of amino groups, greatly promoting nitrogen incorporation into organic solids. These findings not only shed light on how seasonal variations into Pluto's atmosphere composition influence haze formation pathways, but also provide critical parameters to interpret observational data and to improve photochemical and microphysical models of planetary hazes.

# 1. Introduction

Pluto is a Kuiper Belt dwarf planet orbiting the Sun at 30-49 Au with a 248-year orbital period. It hosts a tenuous nitrogen-dominated atmosphere with minor $CH_4$ and CO, maintained in vapor-pressure equilibrium with surface ices. With an average surface temperature of ~44 K and pressure of ~1 Pa (G. R. Gladstone et al. 2016), the atmosphere undergoes strong seasonal variability driven by Pluto's high obliquity and eccentric orbit (C. B. Olkin et al. 2015; L. A. Young 2013). Because vapor pressure is highly temperature-dependent (C. J. Hansen & D. A. Paige 1996), both surface pressure and atmospheric composition evolve over Pluto's year. Although only ~1/7 of a Pluto year has been observed since the atmosphere's initial observation (J. L. Elliot et al. 1989), ground-based and the New Horizons measurements have revealed changes in gas composition (G. R. Gladstone, et al. 2016; E. Lellouch et al. 2015). Due to differences in volatility, the atmospheric $CH_4$ abundance exhibits significant seasonal variation, while the CO abundance remains relatively stable with an abundance of 515±40 ppm (T. Bertrand & F. Forget 2016; E. Lellouch et al. 2017). Based on observational data (F. Forget et al. 2017; G. R. Gladstone, et al. 2016; E. Lellouch, et al. 2015), modeling of Pluto's atmospheric seasonal evolution reveals that the abundance of $CH_4$ increases from trace amounts during northern summer to ~2.6% in northern winter (T. Bertrand & F. Forget 2016). The dramatic variations in $CH_4$ concentrations are primarily caused by the changes of the total pressure, which is controlled by the equilibrium of nitrogen vapor pressure. The absolute pressure of $CH_4$, however, may not fluctuate as significantly, as it is influenced by the sublimation of seasonal polar frost during the northern and southern spring (T. Bertrand et al. 2019).

Ground-based observations (J. L. Elliot et al. 2003; P. Rannou & G. Durry 2009) and New Horizons data (S. A. Stern et al. 2015) showed that photochemical reactions in Pluto's atmosphere produce haze particles. The UV flux is the main energy driver for photochemical processes in Pluto's atmosphere, and it may vary with Pluto's heliocentric distance due to its highly eccentric orbit. Observations from the New Horizons mission revealed multiple haze layers extending from ~50 to 1000 km in Pluto's atmosphere (G. R. Gladstone, et al. 2016). Current understanding suggests that aerosol formation likely initiates in the upper atmosphere through photochemical processes (P. Gao & K. Ohno 2025). At altitudes above ~400 km, photochemistry driven by the ionization and dissociation of $N_2$, $CH_4$, and CO produces a variety of neutral and ionic species that can polymerize to form complex organic aerosols (M. L. Wong et al. 2017; L. A. Young et al. 2018). Similar high-altitude haze layers have been observed in Titan's atmosphere (R. A. West et al. 2018). Moreover, the Cassini mission detected heavy ions with masses up to ~10,000 amu near ~1000 km altitude (~$10^{-6}$ mbar) (A. J. Coates et al. 2007; V. Vuitton et al. 2009), indicating active molecular growth leading to aerosol precursors. This pressure regime corresponds roughly to altitudes of ~500 to 600 km in Pluto's atmosphere, where photochemical haze formation is also expected. As these particles descend to lower altitudes, they may act as condensation nuclei for volatile species. At ~200 to 400 km, condensation of volatile species, such as $C_2H_2$, $C_2H_4$, $C_2H_6$, and HCN onto existing photochemical aerosol particles is expected to dominate (A. Luspay-Kuti et al. 2017; K. Mandt et al. 2017). Atmospheric modeling further suggests that this region corresponds to the formation of ice-rich aerosol particles (P. Lavvas et al. 2020). Analogous processes occur in Titan's atmosphere, where photochemical products such as $C_6H_6$, HCN, $C_2N_2$, and $HC_3N$, condense to form ice clouds (e.g., F. M. Flasar & R. K. Achterberg 2009). These species together likely contribute to the vertically stratified haze layers observed in Pluto's atmosphere.

As haze particles absorb and scatter light differently than gases, these haze particles significantly influence Pluto's temperature structure and climate (T. Bertrand et al. 2025; X. Zhang et al. 2017). Seasonal variations in gas composition are therefore expected to alter haze formation pathways and composition, yet the underlying processes remain poorly understood. Laboratory simulations are thus essential for probing haze formation mechanisms. This approach has significantly advanced our understanding of haze formation in Titan's atmosphere over decades (M. L. Cable et al. 2012; C. Sagan & B. N. Khare 1979) and has been extended to Pluto (L. Jovanović et al. 2021; L. Jovanović et al. 2020), Triton (G. D. McDonald et al. 1994; S. E. Moran et al. 2022), and even exoplanets (L. Gavilan et al. 2018; C. He et al. 2018; C. He et al. 2020; S. M. Hörst et al. 2018). Laboratory-produced analogs of such atmospheric haze are termed 'tholins' (Sagan & Khare 1979). Although laboratory experiments have investigated haze formation in $N_2$-$CH_4$ atmospheres for decades (M. L. Cable, et al. 2012; M. Nuevo et al. 2022; E. Sciamma-O'Brien et al. 2017), experiments including CO—a significant component of Pluto's atmosphere—have only recently emerged (S. M. Hörst & M. A. Tolbert 2014; L. Jovanović, et al. 2020). These studies investigated gas- and solid-phase products compositions (J. M. Bernard et al. 2003; B. Fleury et al. 2014; C. He et al. 2017; L. Jovanović, et al. 2020; S. E. Moran, et al. 2022; Z. Yang et al. 2025), and optical properties of tholins (M. Fayolle et al. 2021; L. Jovanović, et al. 2021; S. E. Moran, et al. 2022), advancing our understanding of CO's role in photochemical reactions. However, these studies primarily focus on the impact of CO concentration on product types and yields, with less discussion on the influence of seasonal variations on photochemical reactions in Pluto's atmosphere—dominated by $CH_4$ concentration changes.

Therefore, in this work, we simulated Pluto's atmospheric chemistry using a glow discharge. We used initial gas mixtures with fixed CO and varied $CH_4$ ratios to systematically dissect the impact of seasonal change on the physical and chemical properties of haze particles. During experiments, we monitored the gas compositions using a residual gas analyzer (RGA), and characterized the haze particle size, density, yield, and compositions using atomic force microscopy (AFM), a gas pycnometer, Fourier-transform infrared (FTIR) spectroscopy, and very high-resolution mass spectrometry (VHRMS).

## 2. Materials and Methods

We conducted 3 sets of Pluto atmospheric photochemical simulation experiments using the PHAZER setup (C. He, et al. 2017). To reproduce seasonal changes of gas abundances, CO was fixed at 515 ppm and $CH_4$ varied at 0.1%, 0.6%, and 5% in $N_2$. The gases were cooled to 100 K and introduced into the chamber at 5 standard cubic centimeters per minute (sccm). The pressure in the chamber is maintained at 1.5 Torr (~ 200 Pa) during the experiments. Although this pressure is much higher than that of Pluto's upper atmosphere (~$10^{-4}$ Pa at 500 km; (P. Lavvas, et al. 2020), the elevated pressure reduces the molecular mean free path and accelerates reaction kinetics, allowing the chemistry to proceed on experimentally accessible timescales. A glow discharge initiated chemical reactions in the gas mixture.

During experiments, the gas composition was continuously monitored using the RGA (SRS, Inc.), with gases ionized at 70 eV and measured from 1–100 m/z at 0.5 amu resolution. Precise identification and quantification of species from RGA data is challenging due to signal overlap and the limited resolution. To address this, we deconvoluted the mass spectra by using a Monte Carlo-based algorithm (T. Gautier et al. 2020) calibrated with the NIST mass spectral database. This method has been validated in prior studies (J. Bourgalais et al. 2020; C. He et al. 2022; J. Serigano et al. 2020; J. Serigano et al. 2022; S. Wang et al. 2025). In this study, the selection of species for the deconvolution analysis is guided by the known composition of Pluto's upper atmosphere, which is dominated by $N_2$, $CH_4$, and CO. Photochemical reactions involving these primary constituents are expected to produce a range of small molecules composed of C, H, O, and N elements. In our experiments, the strongest mass spectral signals were observed primarily below m/z = 50. According to the NIST mass spectral database, only 49 stable CHON-containing gas species can occur within this mass range. Therefore, all 49 candidate species were included in the initial deconvolution procedure to ensure that all chemically plausible contributors were considered. The results showed that only 22 species contributed signals significantly above the background level. These 22 species were therefore retained for the final fitting procedure, reducing unnecessary parameters while improving the robustness of the spectral reconstruction.

After the experiments, solid products were collected in a dry-$N_2$ glovebox and weighed using an analytical balance (Sartorius) to determine haze yield. Particle density was then measured with an AccuPyc II 1340 gas pycnometer (Micromeritics), following procedures described previously (C. He et al. 2024). Smooth mica substrates were pre-positioned inside the chamber before experiments for subsequent atomic force microscopy (AFM) measurements. A Bruker Dimension 3100 AFM equipped with ultrasharp silicon probes (SHR150, BudgetSensors; tip radius < 1 nm; cone angle < 20°) was employed for high-resolution imaging. The equipped ultrasharp silicon probe enables the AFM imaging with lateral resolution <1 nm. Measurements were conducted in tapping mode, consistent with established methodologies (C. He, et al. 2017; C. He, et al. 2024). The particle size distribution and mean diameter were determined via AFM image analysis. For experiments with very low $CH_4$ mixing ratios, where insufficient solid product was available for collection and weighing, the yields were estimated based on the number and size of particles from AFM analysis.

The compositions of solid samples were characterized by FTIR and VHRMS. Transmission spectra (4000–500 $cm^{-1}$, equivalent to 2.5–20 μm) of tholin-KBr pellets were acquired using a Vertex 70v spectrometer (Bruker Optics) equipped with a KBr beamsplitter and a DLaTGS detector under vacuum (< 0.2 mbar), enabling identification of functional groups. The FTIR measurement procedure followed established protocols (C. He, et al. 2024). First, we prepared a pure KBr pellet and tholin-KBr pellets at a specific ratio (~0.2%) using a digital press (CrushIR 15 Ton Digital Press, Pike Technologies). For each FTIR measurement, 500 scans were acquired with a resolution of ~0.5 $cm^{-1}$, and the total acquisition time was about 5 minutes. Additionally, the transmittance of a pure KBr pellet was measured as a reference. The transmittance spectrum of the sample was then calculated as the ratio of the sample measurement to the reference measurement. To analyze the detailed chemical composition of the haze analog samples, we utilized very high-resolution mass spectrometry with a Thermo Fisher Scientific LTQ-Orbitrap XL mass spectrometer equipped with an Ion Max electrospray ionization source (ESI; IPAG, Grenoble, FR). The instrument offers a resolving power (m/Δm) of at least $10^5$ up to m/z 300 and an exact mass accuracy of ±2 ppm. The Pluto haze analog was dissolved in methanol ($CH_3OH$) at a concentration of 1 mg/mL. The mixture was sonicated for 1 hour, followed by centrifugation at 10,000 rpm for 10 minutes. Notably, the 0.1% and 0.6% $CH_4$ samples are only partially soluble in methanol, whereas the majority of the 5% $CH_4$ sample dissolves. The resulting soluble fraction was further diluted in methanol to 1 mg/mL and injected into the Orbitrap via electrospray ionization (ESI), obtaining mass-to-charge (m/z) measurements in the range of 50–300. Mass spectra of blank references were acquired with the Orbitrap under the same settings and were used to reduce noise in the sample data. This procedure ensured that background peaks did not interfere with the identification of molecular formulas in the samples. The sample preparation and data processing followed previous studies (S. E. Moran et al. 2020). For molecular formula assignment, peaks were matched to possible combinations of C, H, O, and N atoms ($C_xH_yO_zN_w$, where x, y, z, w ≥ 0 and integers). To avoid unlikely formulas, following rules were applied: 1/2 < H/C < 2, N/C < 4/3, and O/C < 3/2. Additionally, candidate formulas with a mass error greater than 0.0005 Da relative to the measured peak were excluded. From the identified molecular formulas, we calculated the average molecular formulas, unsaturation degree, and elemental ratios to infer potential formation pathways.

## 3. Results and Discussion

### 3.1. Gas-phase products

The gas-phase mass spectrometry reveals the evolution of gaseous species under glow discharge. To identify peaks with significant intensity changes before and during the experiment, we normalized the mass spectra (MS of the original gas mixture without plasma and MS of the gas mixture when the plasma is on) relative to the intensity of 28 m/z of the original gas mixture (which is $1.13 \times 10^{-5}$ Torr). The peak 28 m/z is the dominant peak and its peak intensity remains stable enough to serve as a normalizing reference. Figure 1a shows the deconvoluted mass spectrum from the 5% $CH_4$ experiment as an example. By comparing the background gas with the plasma experimental gas, we identified the newly formed gaseous products. Figure 1b shows the main gas products in the three experiments (0.1%, 0.6%, and 5% $CH_4$). Across the three experiments, HCN is the most abundant product, which has been detected in previous experiments simulating $N_2/CH_4$ atmospheres (C. He, et al. 2017; Z. Perrin et al. 2021). $C_2H_4O$ is also detected, consistent with previous studies investigating the effect of CO on the $N_2/CH_4$ system (J. M. Bernard, et al. 2003). These detections suggest that similar chemical processes may regulate the chemistry in $N_2/CH_4$ atmospheres such as those of Titan and Pluto. A comparison of the three experimental runs shows that the variety of the gas products increases significantly with $CH_4$ ratio, which is expected as $CH_4$ is relatively easily excited or ionized and serves as a better carbon source in photochemical reactions. The gas species can further react to produce more complex molecules and ultimately solid hazes. The results indicate that initial $CH_4$ concentration impacts the production rate and formation pathways of haze particles in our experiments.

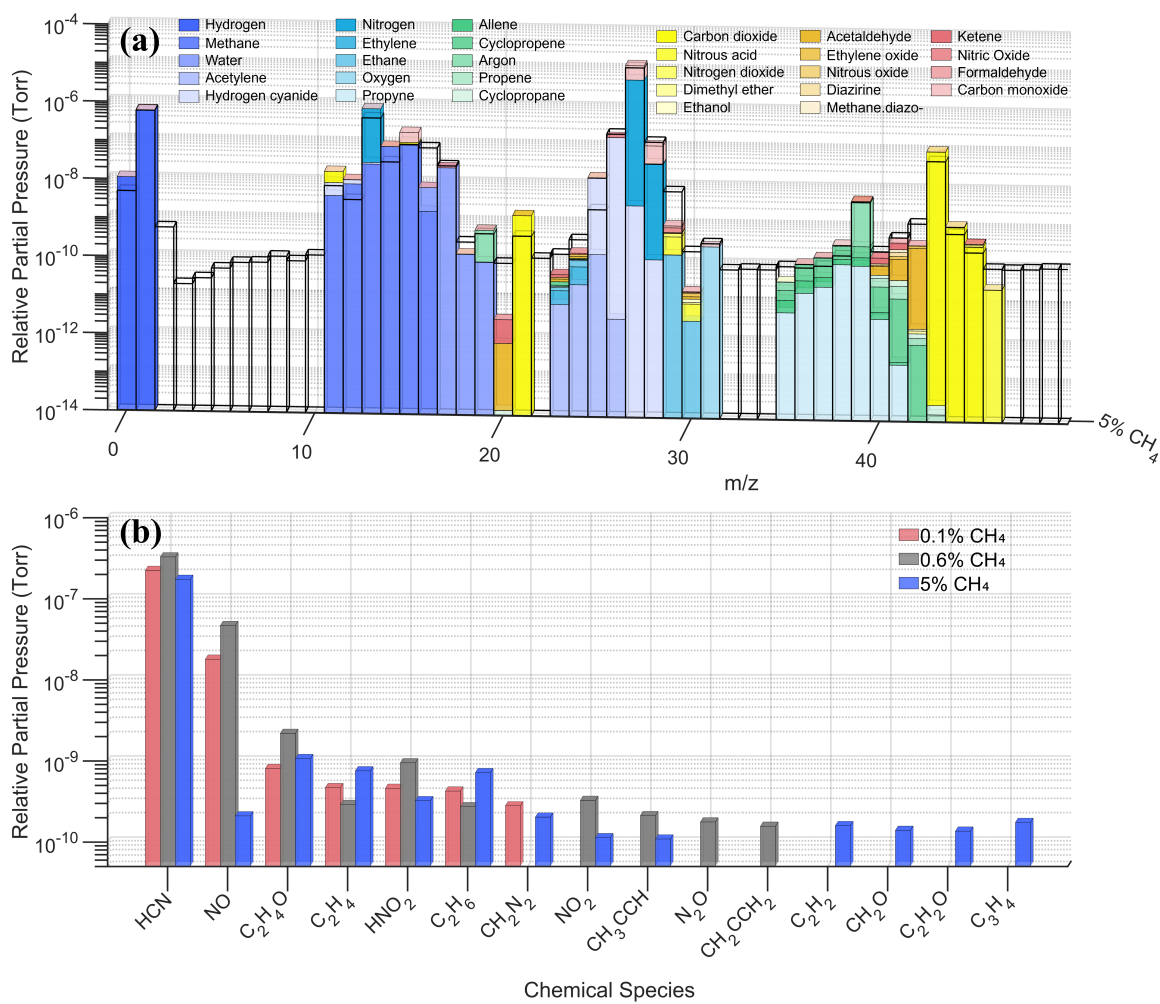

Figure 1. The mass spectrum deconvolution results of the 5% CH$_4$ experiment (a): the black outline represents the normalized RGA data, and the colored segments indicate the contribution of different species to the mass as calculated by the model. The signal intensity of the RGA correlates with the partial pressure of the detected gases. The lower panel (b) shows the deconvolution results in the three experiments with different CH$_4$ mixing ratios (0.1%, 0.6%, and 5%). The major products are defined as species with relative partial pressure greater than 1×10$^{-10}$, while species with partial pressure below 1×10$^{-10}$ are considered to be indistinguishable from noise and do not represent the primary pathways or trends for product formation, thus are not listed here.

### 3.2. Density and Production Rate of Haze Particles

For the 5% CH$_4$ experiment, the density of the collected solid particles was measured as 1.35 g/cm$^3$ using a gas pycnometer, with 0.26% uncertainty from 20 repeats. This value is comparable to previously reported densities (1.3–1.4 g/cm$^3$) for laboratory-produced tholin particles under a range of N$_2$-CH$_4$ atmospheric conditions (e.g., C. He, et al. 2017; H. Imanaka et al. 2012). Imanaka et al. (2012) measured the density of tholins produced from 10% CH$_4$ in N$_2$ at pressures of 1.6 and 23 mbar using a gas pycnometer and found densities of 1.31-1.38 g/cm$^3$. He et al. (2017) reported that the density varies with the CO mixing ratio, but remains relatively stable (1.34 to 1.35 g/cm$^3$) in the low-CO regime (0–0.05%). The lower CH$_4$ cases did not produce enough solid samples for pycnometer density measurement, but their densities are expected to fall within the range of 1.2-1.5 g/cm$^3$. Particle density affects particle sedimentation and aggregation, impacting vertical distribution of haze particles. This influences temperature, climate, and spectral observations of Pluto's atmosphere (E. Lellouch et al. 2025). Therefore, the measured density is useful for microphysical modeling and observation analysis of Pluto (S. Chen et al. 2024; P. Gao et al. 2017).

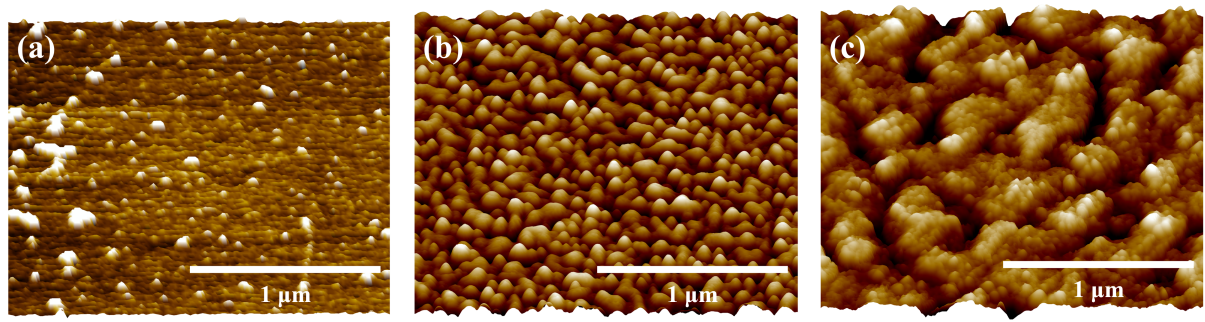

Figure 2. AFM images of three film samples produced with different $CH_4$ concentrations in the initial gases: (a) 0.1% $CH_4$, (b) 0.6% $CH_4$, (c) 5% $CH_4$.

The yield of haze particles varies significantly with the methane concentration. At lower $CH_4$ ratios, the solid production is too small to collect directly, so yields were estimated from AFM-derived particle size and distribution. From image analysis and calculations, we found that the average particle diameter in the 0.1% $CH_4$ mixture was 42±1 nm (10 to 110 nm range, 73.9% within 20-60 nm range; Figure 2a), while for the 0.6% $CH_4$ group, it was 44±1 nm (10 to 90 nm range, 74.7% within 20-60 nm range; Figure 2b). Using the data, we estimated the volume of the particles based on their size, and combined it with the density data to calculate the yield of tholins. We assume that particles are uniformly distributed on the chamber walls, and that only a monolayer of tholins particles is deposited on the mica substrate; therefore, the estimated yield is a lower limit. The estimated yields are $1.26 \times 10^{-6}$ g/h (0.1% $CH_4$) and $4.04 \times 10^{-4}$ g/h (0.6% $CH_4$). For the 0.1% and 0.6% $CH_4$ samples, the AFM images predominantly show dispersed, minimally overlapping particles that appear to be monomers, without obvious large-scale aggregation or multilayer stacking. Therefore, we assume that particles are uniformly distributed on the chamber walls, and that only a monolayer of tholins particles is deposited on the mica substrate. However, we cannot completely rule out the possibility of multilayer formation. As a result, the AFM-estimated yield should be considered as a lower limit. Nonetheless, this approximation provides a reasonable constraint on the overall haze production trend under low-yield conditions, as demonstrated in previous studies (C. He, et al. 2018; C. He, et al. 2020).

For the 5% $CH_4$ sample, the AFM images show the particle size distribution ranges from 10 to 90 nm, with 93.6% concentrated in 20-60 nm and an average diameter of 31±1 nm. The estimated yield is $1.04 \times 10^{-5}$ g/h based on the size distribution. For this sample, enough solid material was collected and weighed. From the total collected mass and the reaction duration, the actual haze production rate was calculated to be $6.21 \times 10^{-3}$ g/h. The AFM-estimated yield is more than two orders of magnitude lower because multiple haze layers are formed in this experiment while the AFM analysis only accounts for the top layer of the deposited particles. Moreover, Figure 2c shows that the smaller monomers have already aggregated to larger particles, approaching 200 nm in size. This behavior is consistent with prior observations of similarly large overlapping particles with scanning electron microscopy (M. Fayolle, et al. 2021).

3.3 FTIR Spectra of Haze Particles

Figure 3a presents the FTIR transmission spectra of tholins produced under different $CH_4$ concentrations, showing the absorption characteristics of various functional groups including $-NH_2$ (3475-3350 $cm^{-1}$), O-H (3500–3200 $cm^{-1}$), $-C\equiv N/-N\equiv C$ (2270-2100 $cm^{-1}$), $C\equiv C$ (2250-2100 $cm^{-1}$), -N=C=N- (2152-2128 $cm^{-1}$), -C=C=N (~2018 $cm^{-1}$), C=O (1740-1720 $cm^{-1}$), C=C (1680-1630 $cm^{-1}$), -N=O (1460-1425 $cm^{-1}$), $-CH_3$ (1338 or 1384 $cm^{-1}$), and $\equiv$C-H (610–681 $cm^{-1}$) (P. R. Griffiths 1992). These functional groups were also observed in the 1% $CH_4$ Pluto tholin by Jovanović et al. (2020) using the Attenuated Total Reflectance technique, suggesting that similar chemical processes occur under different experimental setups with comparable initial gas compositions. For the 0.1% $CH_4$ tholin, the overall absorptions are relatively weaker. As the $CH_4$ ratio increases to 0.6%, the O-H peak noticeably weakens, while the $-C\equiv N/-N\equiv C$ and $C\equiv C$ peaks (2270–2100 $cm^{-1}$) intensify significantly. The 5% $CH_4$ tholin exhibits remarkable spectral differences. The N-H absorption at 3400–3300 $cm^{-1}$ is much stronger with a slight redshift, indicating increased N–H groups and the formation of hydrogen bonds in the sample. However, the N-H and O-H stretching bands partially overlap with each other, so it is difficult to compare the trend of O-H groups changing with $CH_4$ concentration. The -C=C=N feature (~2018 $cm^{-1}$) nearly disappears, indicating reduced unsaturation of the carbon backbone. More prominent absorption peaks appear between 1200 and 1700 $cm^{-1}$, contributed by benzene ring vibration (1600-1500 $cm^{-1}$), N=N (~1550 $cm^{-1}$), and $-NO_2$ (1621-1550 $cm^{-1}$) groups, indicating the formation of more complex functional groups in tholins. In contrast, the $C\equiv C$ absorption at 700–600 $cm^{-1}$ is notably weak.

The distinct differences in functional group types and peak intensities across the three experimental conditions indicate that the formation pathways of tholins vary with $CH_4$ mixing ratios, particularly regarding the way that nitrogen is incorporated in organic products. Low $CH_4$ concentration in the initial gas mixture means the low amount of hydrogen available in the system, limiting the formation of amine groups. For the low $CH_4$ cases, lower amine signature (3000-3500 $cm^{-1}$, Figure 3a) does not necessarily indicate lower nitrogen content in the sample. The nitrogen could still incorporate as C=N, $C\equiv N$, C=C=N, or N=C=N functional groups. However, the vibration modes of $C\equiv N$, C=C=N and N=C=N overlap with $C\equiv C$ in the 2000-2200 $cm^{-1}$ region, so it is difficult to determine the relative contribution of each group. Although the infrared spectrum of the 5% $CH_4$ tholin differs notably from those at the lower $CH_4$ ratios, it closely resembles that of Titan tholins in the characteristic functional groups, likely reflecting the comparable initial gas compositions (Z. Yang, et al. 2025). As shown in Figure 3b, under the same $CH_4$ ratio but with different CO ratios, the types of functional groups in the Titan's haze analogs are similar to those in Pluto, with only certain peaks experiencing small redshift or blueshift. This further suggests that, under similar initial gas conditions in this experiment, the photochemical reaction may have similarities in pathways, leading to the formation of similar types of functional groups in the solid particles. However, differences in the polarity or ratio of certain substances result in shifts in the infrared spectra. The compositional details of the Pluto tholins were further analyzed using VHRMS, as described below.

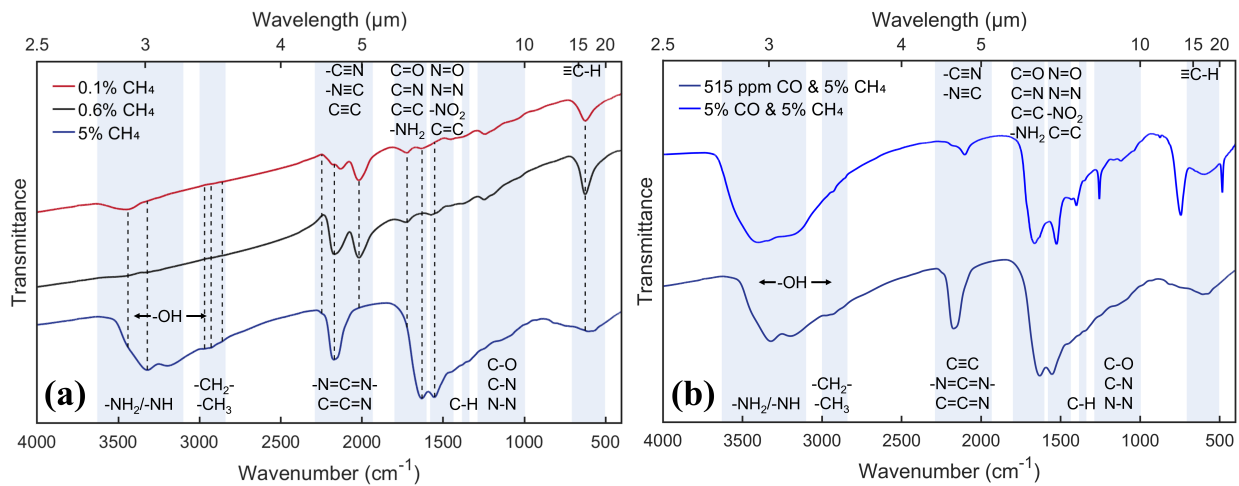

Figure 3. (a) FTIR spectra of three tholins from 4000 to 500 cm$^{-1}$, where the red, black, and blue lines represent the 0.1%, 0.6%, and 5% CH$_4$ tholins, respectively; (b) Comparison of 5% CH$_4$ tholin in this study with Titan tholins in previous work (90% N$_2$/5% CH$_4$/ 5% CO, Z. Yang, et al. 2025). The corresponding functional groups are marked near their absorption features.

### 3.4. Very High-Resolution Mass Spectrometry of Haze Particles

Figure 4 shows the very high-resolution mass spectra of tholins (positive and negative ionization modes) from 50–300 amu under different CH$_4$ mixing ratios. The peak intensities of the 0.1% and 0.6% CH$_4$ tholins are relatively low but still about five times those of the blank references, whereas the 5% CH$_4$ tholin exhibits much stronger signals, roughly 25 times greater. For all three tholins, the peaks in positive mode are more intense than those in negative mode. The weaker signals of the 0.1% and 0.6% CH$_4$ tholins are likely due to their lower yields and poor solubility in methanol, resulting in spectra with less regularity. In contrast, the 5% CH$_4$ tholin displays a repeating pattern near ~14 amu in both positive and negative modes, which is characteristic of organic homologues. Thousands detected mass peaks confirm that these tholins contain complex organic molecules.

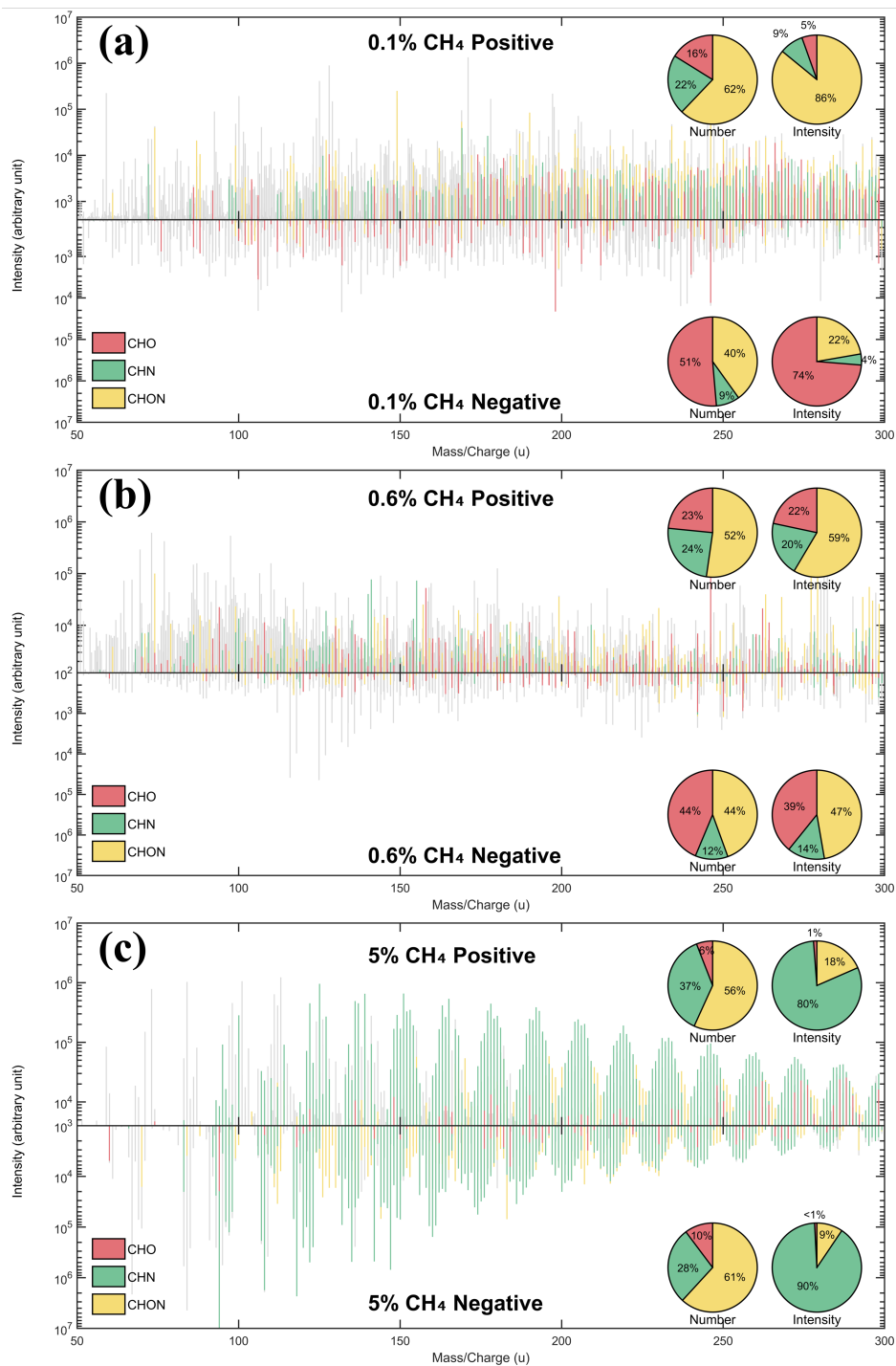

Figure 4. Very high-resolution mass spectra of tholins produced with different CH₄ mixing ratios ((a) 0.1%, (b) 0.6%, and (c) 5%) in both positive and negative ion mode. The red, green, and yellow colors correspond to the CHO, CHN, and CHON subgroups, respectively. Additionally, gray represents data that could not match any molecular formula within the allowable mass precision (<0.0001). The two pie charts in the upper-right corner of each subplot illustrate (i) the number percentage of molecular formulas within each subgroup, and (ii) the abundance percentage based on the summed peak intensities of matched organic molecules.

Based on the exact mass, the detected peaks were assigned to unique molecular formulas, which vary in positive and negative modes of the three tholins. In negative mode, the numbers of matched formulas are 294, 389, and 2127 for the 0.1%, 0.6%, and 5% $CH_4$ tholins, respectively; while the numbers increase to 1370, 1994, and 4147 in the positive mode. More organic formulas are detected with the positive ion mode for all three tholins, suggesting the samples possess more basic characteristics (easily ionized in positive mode via making adducts with proton) than acidic functional groups (forming negative ions by losing proton). Moreover, increasing $CH_4$ mixing ratio significantly raises both the number of matched formulas and their peak intensities. The total peak intensities increase from $8.6 \times 10^6$ (0.1% $CH_4$) to $1.0 \times 10^7$ (0.6% $CH_4$) and $1.3 \times 10^8$ (5% $CH_4$), consistent with higher solubility and enhanced chemical complexity of the solid products at higher $CH_4$ ratio. Based on the elemental composition, all matched formulas fall into CHO, CHN, and CHON groups. CHON compounds dominate the assigned formulas in all samples. With increasing $CH_4$ ratio, the proportion of CHON-organics decreases, while the intensities of CHN-organics increase markedly. In negative mode, CHO species are more abundant and diverse than in positive mode, a typical trend for oxygen-containing molecules (S. E. Moran, et al. 2020).

Based on the matched molecular formulas and their intensities, we calculated the average molecular formulas for each sample, which are $C_{12.33} H_{18.23} O_{1.53} N_{2.15}$, $C_{11.02} H_{18.29} O_{1.68} N_{2.01}$, and $C_{7.50} H_{12.13} O_{1.34} N_{4.42}$ for the 0.1%, 0.6%, and 5% $CH_4$ tholins, respectively. The average molecular formulas were determined based on both positive and negative ion modes, and the abundances from each mode were appropriately weighted and combined. The higher nitrogen content in the 5% $CH_4$ products indicates that increasing $CH_4$ promotes nitrogen incorporation into the solid products, forming nitrogen-containing functional groups (like -C≡N/-N≡C and -N-H as shown by FTIR). Meanwhile, the reduced carbon number reflects the larger abundance and stronger signals of m/z 100–200 species in the 5% $CH_4$ tholin (Figure 4c).

To visualize the compositional characteristic of identified formulas, we plotted the Van Krevelen diagrams of the three samples (Figure 5), where the circle position reflects the atomic ratios (H/C, O/C and N/C) of the matched molecular formulas, the size of each circle represents the intensity of the matched molecular formula peak in VHRMS, the color indicates the double bond equivalent (DBE) calculated based on the molecular formulas, while the spatial distribution reflects compositional relationships between molecules. As shown in Figure 5, the 5% $CH_4$ tholin exhibits a much greater variety of matched molecular formulas than the lower $CH_4$ tholins. The distributions of elemental ratios are relatively dispersed in all cases. Under the same H/C ratio and degree of unsaturation, the N/C ratios are more scattered than the O/C ratios, especially for highly unsaturated species (low H/C) and the 5% $CH_4$ sample. This indicates that the higher degree of unsaturation in these molecules more likely originates from C=C, -C≡C-, or -C≡N, rather than from oxygen-containing groups like C=O.

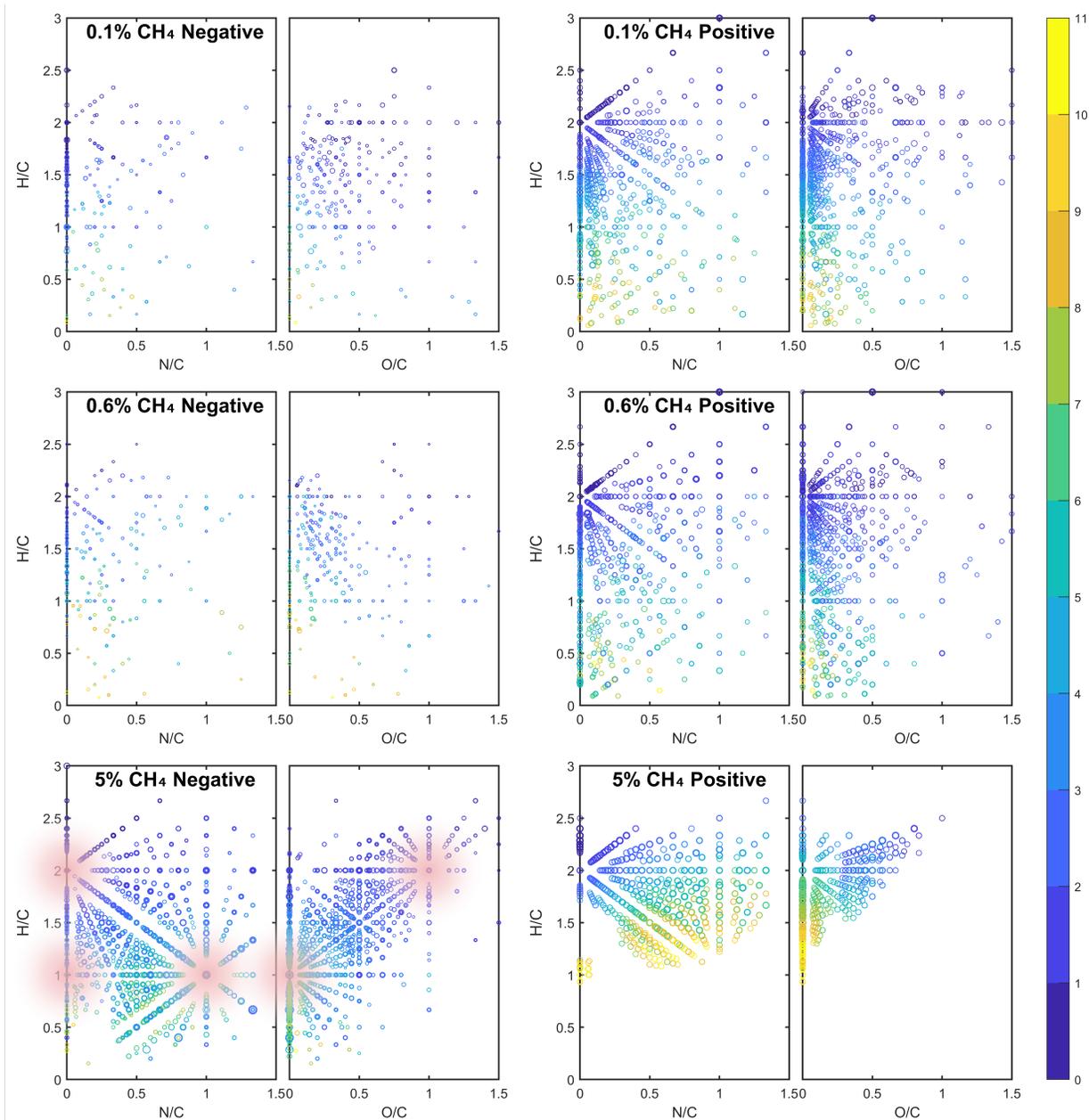

Figure 5. Van Krevelen diagrams (H/C vs. O/C and H/C vs. N/C) of the positive and negative ionization modes at the three experimental conditions. The color code represents the molecules' DBEs, and the symbol size is proportional to the peak intensity of the VHRMS. The centers of the radiating patterns are indicated by semi-transparent red dots in the lower-left panels.

Additionally, for the 0.6% $CH_4$ and 5% $CH_4$ samples, several distinct radiation centers appear in the Van Krevelen diagrams at (0, 1), (0, 2), and (1, 1) in the H/C vs N/C diagram, and (0, 1) and (1, 2) in the H/C vs O/C diagram. Similar patterns have been observed and thoroughly discussed in Wang et al. (2025), suggesting potential reaction precursors like $C_2H_2$, $C_2H_4$, HCN, and $CH_2O$, during the atmospheric chemical processes. The radiation centers are defined based on the statistical distribution of assigned molecular formulas, rather than solely on peak intensities. The purpose of these centers is to identify potential precursor units or repeating motifs involved in haze formation. Because the reaction pathways are highly complex, products likely arise from multiple polymerization and combination reactions of small molecules. Thus, radiation centers represent statistical hubs around which related formulas cluster, rather than necessarily corresponding to the most intense peaks. In this context, molecules near a radiation center may reflect structures formed through repeated incorporation of a specific precursor, even if the largest peaks are not located exactly at the center. These precursors undergo polymerization, co-polymerization, and incorporation reactions, producing molecules that form the observed radiating patterns. In the H/C < 1 region, all three samples exhibit highly unsaturated molecular formulas with higher nitrogen and oxygen contents, but no radiating structures. This suggests that these molecules are not dominated by the polymerization of one specific precursor, but more likely contain functional groups such as N=N and N=O, consistent with the FTIR results.

The Orbitrap results show that the C/N and C/O ratios in the 5% $CH_4$ sample are lower than those in the 0.1% and 0.6% $CH_4$ samples, despite the higher methane concentration in the initial gas mixture. However, these ratios should not be directly compared among the three samples because the mass spectrometry measurements only reflect the fraction of the material that is soluble in methanol. The FTIR results indicate that the 0.1% and 0.6% $CH_4$ samples have very different chemical nature from the 5% $CH_4$ sample, leading to markedly different solubilities in methanol. The 5% $CH_4$ sample dissolves readily in methanol (>90%), suggesting that the Orbitrap data for this sample likely represent a large fraction of the bulk material. In contrast, the 0.1% and 0.6% $CH_4$ samples show much lower yields and poor solubility, meaning the mass spectrometry results likely represent only a small soluble fraction rather than the bulk composition. This interpretation is consistent with the FTIR results. The 5% $CH_4$ sample shows strong absorption features associated with polar functional groups such as O–H and N–H, indicating the dominance of polar compounds that are readily soluble in methanol. FTIR suggests that the lower-$CH_4$ samples are hydrogen-poor and dominated by unsaturated functional groups, whereas the soluble fraction detected by Orbitrap appears relatively hydrogen-rich. This indicates that only a small subset of polar molecules dissolves in methanol, and thus the derived C/N and C/O ratios are not representative of the bulk composition of these samples.

The same VHRMS technique was also applied to analyze the Pluto tholins from Jovanović et al. (2020), although their measurements were conducted only in positive mode (ESI+/Orbitrap). They examined how the initial CO and $CH_4$ concentrations impact the formation of oxygen-containing molecules in the resulting tholins. Their $P_{600}$ tholin (500 ppm CO and 5% $CH_4$ in $N_2$) was generated from an initial gas mixture effectively equivalent to our 5% $CH_4$ tholin (515 ppm CO and 5% $CH_4$ in $N_2$). Consistent with the similarity in gas composition, the VHRMS results exhibit similar regularity in the ESI+/Orbitrap spectra and comparable distribution in the Van Krevelen diagrams. In contrast, their $P_{400}$ and $P_{CO-free}$ tholins were produced with gas mixtures containing 1% $CH_4$, higher than 0.1% and 0.6% $CH_4$ used in our experiments, and therefore are not directly comparable to our low-$CH_4$ tholins. Nonetheless, their VHRMS results showed that a lower $CH_4$ ratio enhances the incorporation of oxygen and nitrogen into the molecules constituting the Pluto tholins (L. Jovanović, et al. 2020). However, we found that increasing the $CH_4$ ratio promotes the formation of nitrogenous molecules in the resulting tholins. The $CH_4$ ratio impacts the chemistry differently in these two studies, likely reflecting the differences of the experimental setups and conditions (e.g., temperatures, energy type, and flux density). Notably, their experiments were conducted at 300 K, while ours were performed at 100 K—closer to the observed temperature in Pluto's atmosphere.

### 3.5. Discussion

In this study, we investigate the impact of seasonal variations in CH$_4$ content on Pluto's atmospheric photochemistry, particularly the haze formation. The gas-phase composition shows that both the variety and molecular complexity of gas products increase with the CH$_4$ mixing ratio. Some of these gases are also observed in the atmospheres of Titan and Pluto (R. Courtin et al. 2011; E. Lellouch, et al. 2017). A greater diversity of gas products at higher CH$_4$ ratios indicates that more complex photochemical reactions occur within the chemical system, leading to the higher yield of the haze particles. Not only the yield, but also the chemical complexity of haze enhances with CH$_4$ ratio, as demonstrated by the VHRMS data.

The CH$_4$ mixing ratio in Pluto's atmosphere experiences seasonal variation, likely ranging from ~0.1% at aphelion up to a few percent at perihelion. Our study indicates that higher methane ratio increases haze production rates. However, Pluto's highly eccentric orbit introduces coupled seasonal variations in multiple atmospheric parameters, including temperature, total pressure, methane abundance, and UV flux. In particular, the UV flux—responsible for initiating photochemistry in the upper atmosphere—can increase by a factor of ~3 from aphelion to perihelion, enhancing haze production in Pluto's atmosphere. These factors interact in a complex manner to control haze formation and are difficult to reproduce simultaneously under laboratory conditions. In this study, we isolated the effect of CH$_4$ mixing ratio as a controlled variable to investigate its role in haze formation. While this approach provides insight into methane-driven chemistry, it does not fully capture the coupled seasonal variability in Pluto's atmosphere.

Seasonal context further illustrates this complexity. Based on climate model results (T. Bertrand & F. Forget 2016; T. Bertrand, et al. 2019), enhanced solar heating near perihelion drives sublimation of surface ices, increasing atmospheric pressure and reaching a maximum during northern spring, as observed during the New Horizons flyby. Elevated pressure reduces the molecular mean free path and accelerates reaction kinetics, favoring haze production. However, at that time, CH$_4$ abundance were at moderate levels, potentially limiting overall photochemical efficiency. In contrast, at Northern autumn and winter, lower atmospheric pressure may suppress haze formation, whereas higher CH$_4$ abundance could enhance photochemistry and shift haze production to lower altitudes where collisional processes remain effective. Overall, the interplay between orbital forcing and atmospheric chemistry introduces significant complexity in haze formation on Pluto, highlighting the need for further studies to better constrain the seasonal evolution of haze properties.

Similarly, the haze analogues produced in our study may not fully represent the true properties of Pluto's haze, particularly in terms of particle size and morphology. However, AFM images show that the particle sizes of the three haze samples are centered at 20-60 nm, which are consistent with the monomer sizes derived from the New Horizons' spectral inversions (S. Fan et al. 2022; P. Gao, et al. 2017; J. Wang et al. 2023). In addition, the higher haze production rate at the 5% CH$_4$ condition leads to the aggregation of the monomers, forming larger particles ~200 nm. However, these larger particles are densely compacted aggregates of monomers, rather than more linear fractal structures as observed in Pluto's atmosphere by the New Horizons spacecraft (S. Fan, et al. 2022). We speculate that high terrestrial gravity may have influenced the particle growth process in the lab setting, resulting in large compact particles. Owing to Pluto's weak gravity, small haze monomer particles are expected to coagulate into loosely bound fractal aggregates in Pluto's atmosphere. Moreover, both the particle size and density affect particle sedimentation, and therefore the vertical distribution of particles. Our measured particle size and density can inform microphysical model and observational data interpretation.

Further, the haze particles interact with light differently from gas molecules. Particles with the size range we measured can scatter blue light effectively and may be responsible for Pluto's blue sky observed by New Horizon spacecraft. Beside scattering, the haze particles also absorb light in different wavelengths at variable level depending on their optical properties, which are determined by their chemical compositions. The FTIR and VHRMS results show that different $CH_4$ ratios lead to the distinct haze chemical compositions, which is consistent with the findings of Jovanović et al. regarding Pluto's tholins. In their study, different $CH_4$ ratios also caused changes in the chemical composition of the solid products, as reflected in the variation of elemental compositions and unsaturation degree (L. Jovanović, et al. 2020). Therefore, haze particles produced in different seasons could have different optical properties, affecting the temperature profile and climate of Pluto, as well as the observed spectra. Future work will involve detailed analysis of the spectra of tholins and the determination of their optical constants, which will aid in the interpretation of Pluto's atmospheric observations.

## 4. Conclusions

In this work, we investigated the influence of seasonal $CH_4$ variations on Pluto's atmospheric photochemistry and haze formation through controlled laboratory simulations. Using gas mixtures with fixed CO (515 ppm) and $CH_4$ ranging from 0.1% to 5%, we systematically examined gas-phase products and solid haze analogs in terms of yield, particle size, density, and chemical compositions. Higher $CH_4$ concentrations lead to increased chemical complexity in the gas phase, providing a greater diversity of precursors for haze formation. Correspondingly, haze particle yields grow with $CH_4$, reflecting more efficient photochemical pathways. FTIR and VHRMS reveal that $CH_4$ abundance strongly influences both the chemical complexity and functional groups of haze particles, promoting nitrogen-rich organics at higher concentrations.

These findings suggest that Pluto's haze formation pathways, particle size distribution, and chemical complexity vary seasonally, controlled by the interplay of $CH_4$ abundance and solar UV flux. Seasonal changes in haze properties are expected to influence Pluto's atmospheric temperature, climate, and spectral signatures. This study demonstrates that $CH_4$-driven seasonal photochemistry is a primary factor controlling the formation and evolution of haze in Pluto's atmosphere, providing critical constraints for interpreting observations and informing atmospheric and climate models.

## Acknowledgments

The authors gratefully acknowledge the support from the National Natural Science Foundation of China (42322407, 42475132), S.E.M. is supported by NASA through NASA Hubble Fellowship grant HST-HF2-51563 awarded by the Space Telescope Science Institute, which is operated by the Association of Universities for Research in Astronomy, Inc., for NASA, under contract NAS5-26555. V.V. acknowledges support from the French National Research Agency in the framework of the "Investissements d'Avenir" program (ANR-15-IDEX-02), through the funding of the Origin of Life project of the Université Grenoble Alpes and the French Space Agency (CNES) undertheir "Exobiologie, Exoplanètes et Protection Planétaire"program.

## References

Bernard, J. M., Coll, P., Coustenis, A., & Raulin, F. 2003, P&SS, 51, 1003
Bertrand, T., & Forget, F. 2016, Natur, 540, 86
Bertrand, T., Forget, F., Umurhan, O. M., Moore, J. M., Young, L. A., et al. 2019, Icar, 329, 148
Bertrand, T., Lellouch, E., Holler, B., Stansberry, J., Wong, I., et al. 2025, NatAs, 9, 1300
Bourgalais, J., Carrasco, N., Vettier, L., Gautier, T., Blanchet, V., et al. 2020, NatSR, 10, 10009
Cable, M. L., Hörst, S. M., Hodyss, R., Beauchamp, P. M., Smith, M. A., et al. 2012, ChRv, 112, 1882
Chen, S., Adams, D., Fan, S., Gao, P., Young, E., et al. 2024, arXiv e-prints, arXiv:2402.00510
Coates, A. J., Crary, F. J., Lewis, G. R., Young, D. T., Waite, J. H., et al. 2007, GeoRL, 34
Courtin, R., Swinyard, B. M., Moreno, R., Fulton, T., Lellouch, E., et al. 2011, Astronomy & Astrophysics, 536
Elliot, J. L., Dunham, E. W., Bosh, A. S., Slivan, S. M., Young, L. A., et al. 1989, Icar, 77, 148
Elliot, J. L., Person, M. J., & Qu, S. 2003, AJ, 126, 1041
Fan, S., Gao, P., Zhang, X., Adams, D. J., Kutsop, N. W., et al. 2022, Nat Commun, 13, 240
Fayolle, M., Quirico, E., Schmitt, B., Jovanovic, L., Gautier, T., et al. 2021, Icar, 367
Flasar, F. M., & Achterberg, R. K. 2009, Philos Trans A Math Phys Eng Sci, 367, 649
Fleury, B., Carrasco, N., Gautier, T., Mahjoub, A., He, J., et al. 2014, Icar, 238, 221
Forget, F., Bertrand, T., Vangvichith, M., Leconte, J., Millour, E., et al. 2017, Icar, 287, 54
Gao, P., Fan, S., Wong, M. L., Liang, M.-C., Shia, R.-L., et al. 2017, Icar, 287, 116
Gao, P., & Ohno, K. 2025, in Triton and Pluto, 6
Gautier, T., Serigano, J., Bourgalais, J., Hörst, S. M., & Trainer, M. G. 2020, RCMS, 34, e8684
Gavilan, L., Carrasco, N., Hoffmann, S. V., Jones, N. C., & Mason, N. J. 2018, ApJ, 861, 110
Gladstone, G. R., Stern, S. A., Ennico, K., Olkin, C. B., Weaver, H. A., et al. 2016, Sci, 351
Griffiths, P. R. 1992, Vibrational Spectroscopy, 4, 121
Hansen, C. J., & Paige, D. A. 1996, Icar, 120, 247
He, C., Hörst, S. M., Lewis, N. K., Yu, X., Moses, J. I., et al. 2018, ApJL, 856
He, C., Hörst, S. M., Lewis, N. K., Yu, X., Moses, J. I., et al. 2020, NatAs, 4, 986
He, C., Hörst, S. M., Riemer, S., Sebree, J. A., Pauley, N., et al. 2017, ApJL, 841
He, C., Radke, M., Moran, S. E., Hörst, S. M., Lewis, N. K., et al. 2024, NatAs, 8, 182
He, C., Serigano, J., Hörst, S. M., Radke, M., & Sebree, J. A. 2022, ACS Earth and Space Chemistry, 6, 2295
Hörst, S. M., He, C., Lewis, N. K., Kempton, E. M. R., Marley, M. S., et al. 2018, NatAs, 2, 303
Hörst, S. M., & Tolbert, M. A. 2014, ApJ, 781
Imanaka, H., Cruikshank, D. P., Khare, B. N., & McKay, C. P. 2012, Icar, 218, 247
Jovanović, L., Gautier, T., Broch, L., Protopapa, S., Bertrand, T., et al. 2021, Icar, 362, 114398
Jovanović, L., Gautier, T., Vuitton, V., Wolters, C., Bourgalais, J., et al. 2020, Icar, 346
Lavvas, P., Lellouch, E., Strobel, D. F., Gurwell, M. A., Cheng, A. F., et al. 2020, NatAs, 5, 289
Lellouch, E., de Bergh, C., Sicardy, B., Forget, F., Vangvichith, M., et al. 2015, Icar, 246, 268
Lellouch, E., Gurwell, M., Butler, B., Fouchet, T., Lavvas, P., et al. 2017, Icar, 286, 289
Lellouch, E., Wong, I., Lavvas, P., Bertrand, T., Villanueva, G., et al. 2025, Astronomy & Astrophysics, 696
Luspay-Kuti, A., Mandt, K., Jessup, K. L., Kammer, J., Hue, V., et al. 2017, Mon Not R Astron Soc, 472, 104
Mandt, K., Luspay-Kuti, A., Hamel, M., Jessup, K. L., Hue, V., et al. 2017, Mon Not R Astron Soc, 472, 118
McDonald, G. D., Thompson, W. R., Heinrich, M., Khare, B. N., & Sagan, C. 1994, Icar, 108, 137
Moran, S. E., Hörst, S. M., He, C., Radke, M. J., Sebree, J. A., et al. 2022, JGRE, 127
Moran, S. E., Hörst, S. M., Vuitton, V., He, C., Lewis, N. K., et al. 2020, PSJ, 1, 17
Nuevo, M., Sciamma-O'Brien, E., Sandford, S. A., Salama, F., Materese, C. K., et al. 2022, Icar, 376
Olkin, C. B., Young, L. A., Borncamp, D., Pickles, A., Sicardy, B., et al. 2015, Icar, 246, 220
Perrin, Z., Carrasco, N., Chatain, A., Jovanovic, L., Vettier, L., et al. 2021, Processes, 9
Rannou, P., & Durry, G. 2009, JGRE, 114
Sagan, C., & Khare, B. N. 1979, Natur, 277, 102
Sciamma-O'Brien, E., Upton, K. T., & Salama, F. 2017, Icar, 289, 214
Serigano, J., Hörst, S. M., He, C., Gautier, T., Yelle, R. V., et al. 2020, JGRE, 125, e2020JE006427
Serigano, J., Hörst, S. M., He, C., Gautier, T., Yelle, R. V., et al. 2022, JGRE, 127
Stern, S. A., Bagenal, F., Ennico, K., Gladstone, G. R., Grundy, W. M., et al. 2015, Sci, 350
Vuitton, V., Lavvas, P., Yelle, R. V., Galand, M., Wellbrock, A., et al. 2009, P&SS, 57, 1558
Wang, J., Fan, S., Liu, C., Natraj, V., Young, L. A., et al. 2023, PSJ, 4
Wang, S., Yang, Z., He, C., Li, H., Liu, Y., et al. 2025, ApJ, 990
West, R. A., Seignovert, B., Rannou, P., Dumont, P., Turtle, E. P., et al. 2018, NatAs, 2, 495
Wong, M. L., Fan, S., Gao, P., Liang, M.-C., Shia, R.-L., et al. 2017, Icar, 287, 110
Yang, Z., Liu, Y., He, C., Yu, P., Jin, R., et al. 2025, PSJ, 6
Young, L. A. 2013, ApJ, 766
Young, L. A., Kammer, J. A., Steffl, A. J., Gladstone, G. R., Summers, M. E., et al. 2018, Icar, 300, 174
Zhang, X., Strobel, D. F., & Imanaka, H. 2017, Natur, 551, 352